\documentclass[useAMS,usenatbib]{mn2e}

\usepackage{graphicx}
\usepackage{amsmath}
\usepackage{amssymb}
\usepackage{color}
\usepackage{subfig}
\usepackage[normalem]{ulem}

\usepackage{hyperref}
\hypersetup{colorlinks=true,linkcolor=blue,citecolor=blue,filecolor=blue,urlcolor=blue}

\newcommand\msun{\, \rm M_\odot}
\newcommand\rsun{\, \rm R_\odot}
\newcommand\kms{\, \rm km\,s^{-1}}
\newcommand\pc{{\, \rm pc}}
\newcommand\kpc{{\, \rm kpc}}

\newcommand\myr{{\, \rm Myr}}
\newcommand\gyr{{\, \rm Gyr}}
\newcommand\au{{\, \rm AU}}

\newcommand\mstar{M_\star}
\newcommand\imbh{M_\mathrm{IMBH}}
\newcommand\fbin{f_\mathrm{bin}}
\newcommand\rh{r_\mathrm{h}}
\newcommand\vesc{v_\mathrm{esc}}

\newcommand\code[1]{{\tt{}#1}}

\newcommand{\MODBH}{{\sf{}BH}}    
\newcommand{\MODBHBIN}{{\sf{}BH-BIN}} 
\newcommand{\MODBIN}{{\sf{}BIN}}   

\title[IMBHs in binary-rich star clusters]{Intermediate-Mass Black Holes in binary-rich star clusters}
\author[Ladislav \v{S}ubr, Giacomo Fragione and J\"org Dabringhausen]{Ladislav \v{S}ubr$^{1}$\thanks{E-mail: subr@sirrah.troja.mff.cuni.cz}, Giacomo Fragione$^{2}$ and J\"org Dabringhausen$^{1}$\\
$^{1}$Astronomical Institute, Charles University, V Holesovickach 2, CZ-180 00 Praha 8, Czech Republic\\
$^{2}$Racah Institute for Physics, The Hebrew University, Jerusalem 91904, Israel}

\begin{document}

\maketitle

\begin{abstract}
There is both theoretical expectation and some observational clues that intermediate mass black holes reside in nuclei of globular clusters. In order to find an independent indicator for their existence, we investigate in this paper how an IMBH manifests itself through its dynamical interaction with a binary rich globular cluster of moderate extension and mass. By means of direct $N$-body integration we follow the dynamical evolution of models of such a system over a time span of $\approx 0.8\,\gyr$ and compare the cases with and without the primordial binaries as well as with and without the IMBH. In accord with previous results, we show that when present the IMBH develops a power-law density cusp of stars around it, regardless of the binary population in the cluster. If, however, binaries are present, their interaction with the IMBH leads to the production of high velocity escapers at a rate of the order of $0.1\,\myr^{-1}$. These stars may contribute to the population of high-velocity stars observed in the Galaxy. Clusters hosting the IMBH together with high number of binaries also form a denser halo of marginally unbound stars than clusters that lack either the IMBH or the rich binary population. Finally, we show that the binary population leads to an increased rate of direct interactions of stars with the IMBH, potentially observable as tidal disruption events.
\end{abstract}


\section{Introduction}
\label{sec:introduction}

Independently of their spin (and possibly charge), black holes are usually divided into three different categories. Supermassive black holes (SMBHs, $M\gtrsim 10^5 \msun$) reside in the centres of galaxies, where they shape the surrounding gas and stellar content \citep{hop06,alex17}. Stellar-mass black holes (SBHs, $10\msun\lesssim M\lesssim 100\msun$) are the final stage of massive stars lifetime and have been recently observed by LIGO detections in SBH-SBH merger events \citep{abb16,abb17,frak18}. However, for the intermediate-size population, the so-called intermediate-mass black holes (IMBHs) ($100\msun\lesssim M\lesssim 10^5 \msun$), there is not yet a definitive accepted proof of their existence. In the Milky Way, two clusters have been claimed to host an IMBH based on dynamical modeling, namely $\omega$~Cen \citep{noy10,jal12,bau17} and 47~Tuc \citep{kiz17}, but other authors \citep{vdMarel10a,vdMarel10b,zoc17,zoc18} have advised caution with this interpretation. However, the recent observation of a tidal disruption event (TDE) consistent with an IMBH of $\sim 5\times 10^4 \msun$ in an off-centre star cluster, at a distance of $\sim 12.5$ kpc from the centre of the host galaxy, has provided a new, maybe final, proof of their existence \citep{lin18}.

Assuming that the observed $M_\mathrm{SMBH}-\sigma$ relation ($\sigma$ is the velocity dispersion) holds also in the range of IMBH masses \citep{mer13}, the ideal place to look for IMBHs is at the centres of globular clusters (GCs). In GCs, the most massive stars may segregate and merge in the core, forming an object in the mass range of an IMBH \citep{fre06}. \citet{por02} suggested that star clusters with small initial half-mass relaxation times may form an IMBH with a mass up to $\sim 0.1$\% of the mass of the entire star cluster, as a consequence of the high stellar collision rates. Recently, \citet{gie15} found slow and fast scenarios of IMBH mass growth: GCs with a large initial cluster concentration have a larger probability of forming an IMBH earlier and faster. Once formed, an IMBH resides at the centre of its host GC \citep{chatt02a,chatt02b}, and interacts with the host cluster stars and compact objects \citep{lei14}. Unfortunately, both the lack of high-resolution data and accurate N-body modeling of star clusters deny definitive kinematic detections of IMBH in the centres of GCs. 

Galactic nuclei may host IMBHs as well. The centre of the Milky Way may host an IMBH in the central parsec and several IMBHs in its nuclear star cluster, possibly delivered by inspiraling star clusters \citep*{por02,fragk18,fragleiginkoc18}. In galactic nuclei in general, the TDE rate may be enhanced due to the presence of an IMBH \citep{che09,che11,fraglei18}.

IMBHs may be illuminated by Gravitational Wave (GW) events. IMBH-SBH binaries may commonly form in the core of GCs and may merge as intermediate mass ratio inspirals (IMRIs), at a rate as large as $\sim$ few $\mathrm{Gpc}^{-3}\, \mathrm{yr}^{-1}$. IMBH-SBH binaries of different masses can be observed by present and upcoming facilities, such as LIGO, the Einstein Telescope, and LISA \citep*{lei14,ama10,fragk18,frlei18}. Besides GW events, TDEs may also take place in the center of a cluster hosting an IMBH \citep*{bau04a,bau04b,fragleiginkoc18}, whose rate ($10^{-6}$-$10^{-3}$ yr$^{-1}$) can be of the same order of the TDEs driven by the host galaxy SMBH. 

Several efforts have been made in modeling GCs with IMBHs through direct $N$-body simulations. \citet{baum05} investigated the radial density profile of stellar systems with a central IMBH. \citet{tren07} and \citet{gill08} used direct N-body simulations to study clusters with an IMBH and $5$-$10$\% primordial binaries. One of the predictions of these models was that clusters hosting an IMBH can develop a cusp of stars in the projected surface brightness profile \citep{baum05,mioc07}, which is usually regarded as an indicator of the presence of an IMBH, even though also other dynamical processes can produce similar signatures \citep[see e.g.][]{hurl07,vesp10}. \citet{lut13} performed $N$-body simulations of GCs hosting IMBHs as a function of the neutron star and SBH retention fraction, while \citet{lei14} studied the co-existence of SBH binaries with an IMBH. Recently, \citet{bau17} has run a large grid of N-body simulations with different mass fractions $M_{\mathrm{IMBH}}/M_{\mathrm{GC}}$ (where $M_{\mathrm{IMBH}}$ and $M_{\mathrm{GC}}$ are the IMBH and GC mass, respectively).

Presence of large number of primordial binaries affects evolution of star clusters, namely their cores during the collapse and subsequent gravo-thermal oscillations \citep[see, e.g.,][for extensive study on this topic]{heggie2006,trenti2007a}. It is also expected that interaction of binaries with an IMBH leads to additional phenomena, e.g., ejection of high velocity stars through the Hills mechanism, similarly to what happens in Galactic Centre \citep{hills88,yut03,brm06,sari2010,brown2018}. In this scenario, a binary star passing close to the IMBH is tidally separated with one component being ejected away at high velocity, while the other one remains tightly bound to the IMBH \citep{frgual18b}.

A full $N$-body treatment of star cluster with primordial binaries and the IMBH is a computationally challenging task and, therefore, only few works on this topic can be found in the literature so far \citep[see, e.g.,][for one of the first attempts]{tren07}.
In this paper, we aim to further fill in this gap, presenting the first direct $N$-body simulations of star clusters hosting an IMBH and a large primordial binary fraction (up to a binary fraction of $50$\%). We use for our simulations a modified version of the \code{NBODY6} code \citep{aarseth03}. We study clusters with or without an IMBH and with different binary fractions, show how the presence of the IMBH affects the central parts of the cluster and discuss the rate at which the IMBH ejects stars out of the cluster into the host galaxy. We then discuss how these ejected stars may contribute to the observed population of hypervelocity (HVSs) and runaway stars (RSs), along with the other mechanisms proposed in literature \citep{hills88,yut03,GPZ2007,sesa2006,zub2013,cap15,fra16,subr16,bou2017,fra17,fgu18,frgual18b}.

The paper is organized as follows. In Section 2, we present the numerical method we use to evolve the GC. In Section 3, we present our results. Finally, in Section 4, we draw our conclusions.

\section{Method}

Our numerical simulations with the \code{NBODY6} code \citep{aarseth03} aim at studying the impact of an IMBH on the properties and the evolution of an aged stellar system like a low-mass GC. The initial structure, stellar populations and binary populations are therefore chosen to represent roughly such a system, while avoiding too many speculative details and keeping the models numerically manageable. IMBHs are thought to form early in the evolution of a massive star cluster (see Section~\ref{sec:introduction}), if they form at all, and their formation is therefore not under scrutiny in the present models. The models are rather set up with a particle representing the IMBH, and compared to models that are set up without an IMBH, but identical parameters otherwise. This allows to identify the impact of the IMBH on such systems, and how the presence of an IMBH may be detected observationally.

The simultaneous treatment of a large population of binaries, as it is likely to be found in GCs (like in any known stellar population), and an IMBH has not been done in previous simulations. The reason is that binary systems are numerically demanding and large differences in the masses of the simulated particles (as they are implied by the presence of an IMBH among stars) are a challenge for direct $N$-body integrators. The simultaneous treatment of these two complications is what limits our simulations to 50000 stars, given that an additional constraint is that for meaningful results on the long-term evolution and stability of these systems, we need to simulate several hundreds of Myr of cluster evolution within reasonable computing time. Thus, 50000 particles sounds like a modest number in comparison to the $10^6$ particles considered by \citet{Wang2016}, but we note that in the simulations of \citet{Wang2016} only $5$\% of the stars are in primordial binaries (in contrast to $50$\% in our simulations), and also the IMBH is missing.

\subsection{Numerical model for a star cluster with binaries and an IMBH}
\label{sec:MODBHBIN}

In the following, we give a detailed description of our setup for the system that is at the centre of attention in the present paper, namely a massive intermediate-age star cluster with a large population of binaries and an IMBH. We will refer to this model as model \MODBHBIN, and name and discuss our choices regarding the stellar mass function (Section~\ref{sec:massfunction}), the initial binary population (Section~\ref{sec:binaries}), and the initial cluster structure (Section~\ref{sec:structure}) in this model. In Section \ref{sec:othermodels}, we will briefly describe some additional $N$-body models that we run in order to evaluate which impact the inclusion of a large binary population and an IMBH has on the evolution of a star cluster.

\begin{table}
\caption{Variable parameters of the models: name, number of stars ($N$), mass of the IMBH ($\imbh$) and primordial binary fraction ($\fbin$).}
\begin{center}
\begin{tabular}{c|ccc}
\hline
 name & $N$ & $\imbh$ & $\fbin$ \\
\hline
\hline
 \MODBHBIN  & $50001$ 		  & $10^3\msun$ & $0.5$  \\
 \MODBH     & $50001$		  & $10^3\msun$ & $0.0$  \\
 \MODBIN    & $50000$ 		  & $0$         & $0.5$  \\
\hline
\end{tabular}
\end{center}
\label{tab:models}
\end{table}

\subsubsection{Mass function}
\label{sec:massfunction}

We assume for model \MODBHBIN \ that the initial masses of stars are distributed according to a broken power-law distribution function with the power-law index $\alpha = -1.3$ for $\mstar < 0.5\msun$ and $\alpha = -2.3$ for $0.5 \msun \le \mstar < m_{\rm max}$. This mass function is known as the canonical stellar initial mass function, and is characteristic for many star formation events, except the most extreme ones \citep{kro01,kro13}.

Furthermore, all stars are assumed to have formed instantaneously in a single event. In reality, stars certainly do not form instantaneously in a star cluster, and some models for explaining the observed element abundances in stars in  GCs even imply that the stars in GCs were born in multiple star formation epochs instead of a single one. However, these events take place on a time scale of several 100 Myrs at most, so that the birth of all stars in an initial star burst becomes a very good approximation when the stellar masses of a several Gyr old GC are considered.

The initial number of particles representing individual stars and stellar remnants is 50000. Of these particles, 44500 represent stars. Their masses are drawn randomly from the canonical IMF in the mass interval $(0.1 \ {\rm M}_{\odot},\, 1 \ {\rm M}_{\odot})$. The upper limit of assumed stellar masses is motivated with the finding that already in a 1 Gyr old stellar population, the most massive surviving stars have a mass below $1\,\msun$, and from then onwards, the upper mass limit for surviving stars keeps on dropping only slowly as the stellar population continues to age (see e.g. the simple stellar population models by \citealt{maraston2005}). Our upper mass limit therefore roughly represents the situation in a few Gyr old star cluster.

The remaining 5500 particles with stellar masses represent stellar remnants; i.e. white dwarfs, neutron stars and stellar-mass black holes. Their number corresponds to the number of stars with a mass above $1\,\msun$ according to the canonical IMF, if the number of stars below $1\,\msun$ is 44500, and the upper mass limit of the IMF, $m_{\rm max}$ is set to $100\,\msun$. The actual upper mass limit of the IMF may be significantly higher (cf. \citealt{Crowther2010}), but if the IMF is canonical, stars with initial masses above 100 $\msun$ are very rare and can therefore be neglected for the purpose of this paper.

The mass attributed to each particle representing a stellar remnant is $1\,\msun$. This is only an approximation, but it seems justifiable, given that the observed masses of white dwarfs in \citet{Kalirai2008} range between $0.5\,\msun$ and $1.1\,\msun$, and given that the observed masses of neutron stars seem all close to a value of $1.35\,\msun$ \citep{Thorsett1999}. Stars with an initial mass above $25\,\msun$ are likely to become stellar-mass black holes which can be much heavier than $1\,\msun$ (compare for example figures~12 and~16 in \citealt{Woosley2002}), but their numbers are suppressed strongly by the steep slope of the canonical IMF at high masses.

We finally include also one additional particle with a mass of $1000\,\msun$, which represents the IMBH. This additional particle is initially placed at a random position in the cluster, but sinks rapidly towards the centre of the cluster through dynamical friction due to its high mass \citep{bin08}.

Thus, in essence, the model \MODBHBIN \ is set up to represent few Gyr old massive star clusters in which stars were formed according to the canonical IMF. Stellar evolution is considered by the choice of the upper mass limit for still luminous stars and the ratio between the number of luminous stars and stellar remnants, while dynamical evolution, which leads to a loss of low-mass stars \citep{Baumgardt2003}, is neglected.

\subsubsection{Binaries}
\label{sec:binaries}

\citet{Kroupa1995} argues that initially nearly every star in a star cluster is part of a binary or a multiple system of even higher order. However, many of these initial binaries are very wide and therefore easily disrupted through encounters. Thus, in dynamically evolved star clusters, the fraction of stars in multiple systems will be much lower.

The actual properties of the binary populations in GCs in not known in much detail, but as an approximation to the binary population in GCs, another population that has undergone strong dynamical processing may serve. If essentially all stars were born in star clusters, as suggested by \citet{Lada2003}, such a population is found in the Galactic field, because the stars in the Galactic field then all come from star clusters that were dissolved through dynamical processes. The properties of binaries in the Galactic field are well studied by \citet{dm91}, which is therefore chosen as the basis for the present work. We note that according to \citet{Minor2013} also the properties of the binary populations in the dwarf Spheroidal Galaxies which he studied are consistent with those \citet{dm91} found for the Galactic field.

The binary fraction, $\fbin$, is defined as the number of binaries over the number of center-of-mass systems, i.e. binaries and single stars in our case. We set $\fbin=0.5$ in our model \MODBHBIN, which is close to the value \citet{dm91} find in the Galactic field (but much lower than $\fbin=1$, which was suggested by \citet{Kroupa1995} for dynamically unevolved star clusters). We note that \citet{Wang2016} assumed $\fbin=0.05$ in their N-Body simulation with $10^6$ individual stars, while \citet{Sollima2007} found $\fbin>0.06$ for the low-density GCs they studied, and in some cases $\fbin \approx 0.5$ seems likely. This motivates our high choice for $\fbin$ in model \MODBHBIN. However, in more massive and thus (for a fixed radius) more dense GCs, the binary fraction appears to be smaller ($0.04 < \fbin < 0.09$, see \citealt{Cool2002} for the case of NGC 6397 and \citealt{Romani1991} for the case of M92; see also \citealt{Sollima2007}), and thus the choice of \citet{Wang2016} seems appropriate for the model they consider.  

The semi-major axes of primordial binaries follow the distribution found by \cite{dm91} for the Galactic field in all models we consider, i.e. a log-normal distribution which peaks at a semi-major axis corresponding to an orbital period of $\sim 180$ yr. The interpretation of this distribution according to \citet{Kroupa1995} is that binaries with very wide orbits, which may have been frequent initially, are already destroyed through dynamical interactions. The initial eccentricities of the binaries are set to zero for simplicity.

The procedure of pairing is based on random number generator. All objects (stars and compact remnants) are first put into a pool ordered according to their mass from the most massive one to the lightest. Starting from the most massive object, we generate a random number which determines whether it will be paired with another one or not (with equal probability for these two cases). If the object is to be a member of a binary, its secondary is selected with uniform probability from the objects of the same type remaining in the pool (i.e., binaries made of main sequence star and compact remnant are not allowed). Both objects are then removed from the pool and the procedure continues on, until all stars are taken from the pool. We do not set up higher-order systems.

\subsubsection{Initial cluster structure}
\label{sec:structure}

In our setup for model \MODBHBIN, the star cluster initially follows the Plummer density profile with half-mass radius, $\rh = 3\pc$. The Plummer density profile has been shown by \citet{Plummer1911} to provide simple, yet very satisfactory fits to the observed density profiles of GCs, and is therefore still a very popular choice for setting up models for star clusters.

The chosen half-mass radius is quite characteristic for GCs in the Milky Way (see figure~8 in \citealt{McLaughlin2000}), but also for extragalactic GCs in the Virgo Cluster \citep{Jordan2005} and the Fornax Cluster \citep{Jordan2015}. The choice for the half-mass radius puts our model, together with the mass attributed by 50000 stars (see Section~\ref{sec:massfunction}), in the regime of low-mass and low-density GCs. This makes the choice of rather high binary fractions suggested in \citet{Sollima2007} for low-density appropriate for them, while the case of a low binary fraction that \citet{Wang2016} use in their models of rather massive GCs is depreciated for our case (see also Section~\ref{sec:binaries}).

The initial positions and velocities were generated using the {\code{plumix}} code \citep{subr08,plumix} which ensures that the IMBH local mass overdensity is somewhat reduced by lower local density of lighter objects.

Finally, all the cluster models are generated in isolation, i.e. without any external tidal field.

\subsection{Other N-body models}
\label{sec:othermodels}

The focus of this work lies on the evolution of massive star clusters with an IMBH \emph{and} binaries, which is the situation depicted with our model {\MODBHBIN}. However, in order to evaluate how the interplay of an IMBH and a large population of primordial binaries affects the evolution of a massive star cluster, we have integrated additional N-body models, which we list below:
\begin{itemize}
\item Model {\MODBIN} is the same as model {\MODBHBIN}, except that there is no particle representing the IMBH included in model {\MODBIN}.
\item  Model {\MODBH} is the same as model {\MODBHBIN}, except that $f_{\rm bin}=0$ in model {\MODBH}, so that there are initially no binaries in the simulation, while they may form dynamically later on.
\end{itemize}

A summary and a quick reference to our models is given in Table~(\ref{tab:models}). For each model, seven realisations which differ in initial positions and velocities of stars, but sharing the overall statistical properties, were integrated. Results presented below are based on averages over the seven realisations of individual models.

\subsection{Numerical integrator}
We used \code{NBODY6} code \citep{aarseth03} for the numerical integration of the equations of motion. We added to the original code routines for logging of beginnings and endings of regularizations into a binary file. We also altered the decision making algorithm for adding particles to neighbor lists. In particular, we weight the standard distance criterion by the mass of the given particle so that the more massive particles are added to the list even when they are at larger distances than the lighter ones. The most prominent target was the IMBH particle which was, due to this modification, a member of the neighbor lists of all other (star) particles. The modification of the neighbor list influences the integration and increases its stability. Among many runtime options of the \code{NBODY6} code, let us specifically mention that we switched off the internal evolution of the stars as well as the post-Newtonian corrections to the stellar dynamics.

\section{Results}

\subsection{Radial density profile}
\begin{figure*}
\begin{center}
\includegraphics[width=0.9\textwidth]{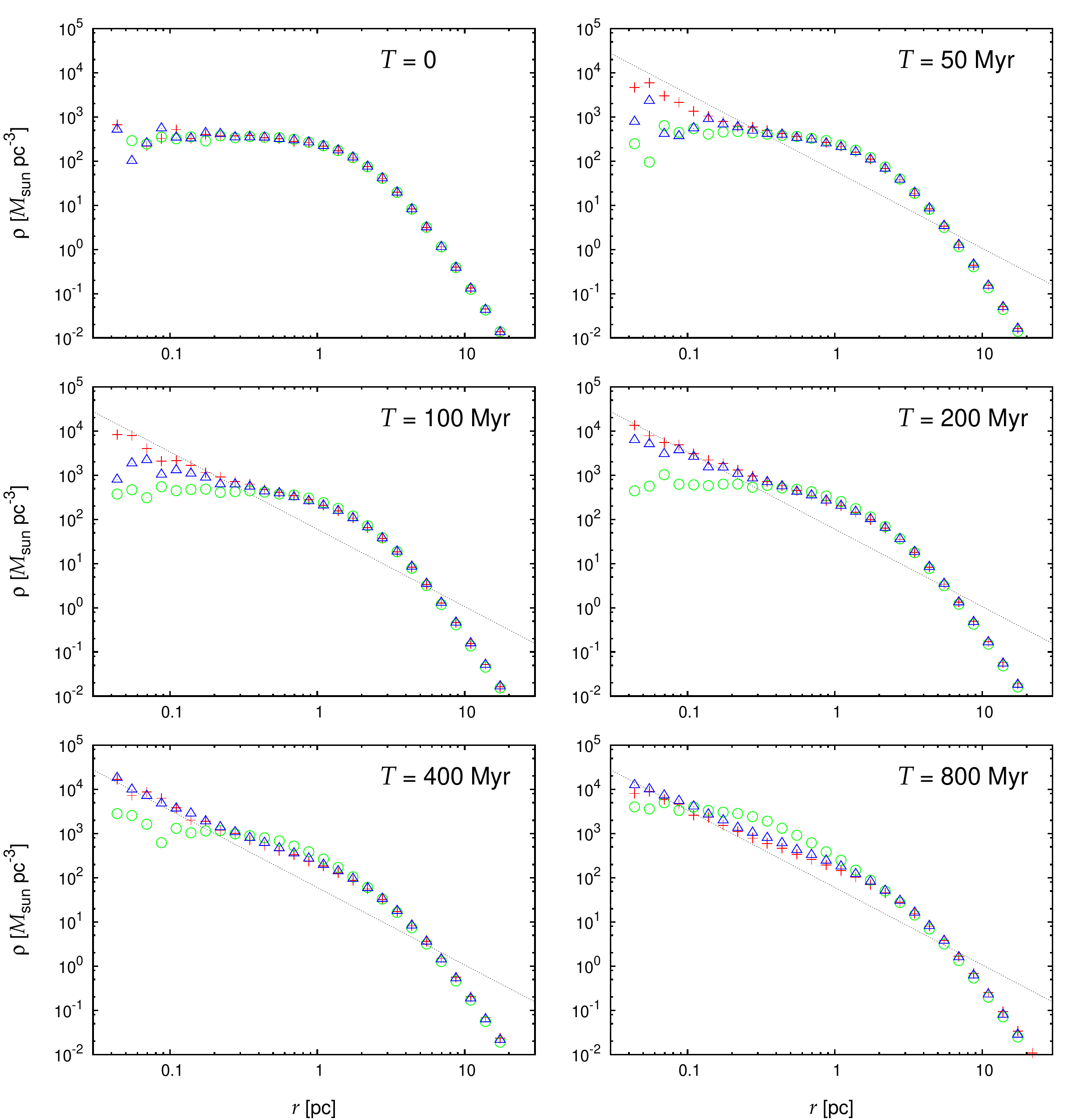}
\end{center}
\caption{\label{fig:rp}
Radial density profiles of models {\MODBHBIN} (red crosses), {\MODBIN} (green circles) and {\MODBH} (blue triangles) at six different epochs (indicated in each panel); dotted line indicates the power-law with slope index $-7/4$.}
\end{figure*}
The IMBH which was set at random position in the star cluster initially, sinks quickly into the cluster centre on the time scale $\lesssim 20\myr$ due to dynamical friction \citep{bin08}. From that time on, a stellar cusp evolves around the IMBH and it is clearly visible in the radial density profile already at $T \geq 100\myr$ (see Fig.~\ref{fig:rp}). The buildup of the cusp is somewhat faster in model {\MODBHBIN} than in {\MODBH}, nevertheless, already at $T \gtrsim 200\myr$, the difference between the two models becomes negligible. We also note that the region below $\approx 0.1\pc$ is occupied by no more than several tens of stars and, therefore, the radial density there fluctuates in time and individual realisations of a particular model may differ from each other. Once the central cusp approaches the equilibriums state, its evolution slows down and both models that include the IMBH share a similar density profile. The central cusp (within the IMBH sphere of influence $r_i$) is well approximated by the \cite{bw76} density profile $\varrho(r) \propto r^{-7/4}$. In agreement with previous works \citep{baum05,mioc07}, we find that a shallow cusp in the projected surface brightness profile is a possible indicator of the presence of an IMBH, but other dynamical processes can produce cuspy profiles as well \citep{hurl07,vesp10}.

The fact that the radial density profile reaches near equilibrium state in the innermost parts of the models including the IMBH, however, does not mean that the central cusp does not lead to any interesting phenomena. At each moment after $T \approx 100\myr$, there are typically one or two stars or stellar remnants orbiting the IMBH with semi-major axis smaller than $\lesssim 10^{-4}\pc$ and eccentricity $e > 0.99$ \citep[see also][for discussion of dynamical formation and properties of close stellar companions to an IMBH]{mac16}. Peak velocities of these stars during passages through the percentres of their orbits are few thousands of kilometers per second, i.e., comparable to velocities of so-called S-stars orbiting the Galactic SMBH \citep[see, e.g.,][]{ghez2005,gillessen2009}. Integration of these stars is in general time demanding due to large accelerations and the fact that the tightly bound system consisting of more than two particles (including the IMBH) is not suitable for standard regularisation techniques implemented in $N$-body integrators. Despite their large binding energy, even the orbits of the most tightly bound stars are being subject to perturbations which often lead to their collision with the IMBH.

The region at distances $10^{-4}\pc \lesssim r \lesssim 0.1\pc$ from the IMBH is typically occupied by several tens of solar-mass objects which can be considered bound to the IMBH. Nevertheless, it appears that the time an individual star spends on the orbit bound to the IMBH is quite short -- more than 60\% of stars are scattered out from this region on the time-scale of $\approx 16\myr$. This quite strong relaxation in the vicinity of the IMBH is very likely another source of computational error which leads to slow down of the integrations.

In our models, we have adopted an initial Plummer profile for the initial density profile of the clusters. Since the cluster loses memory of the initial conditions after roughly one relaxation time, the long-term evolution of the cluster does not depend significantly on the details of the initial conditions, once the IMBH mass and the primordial binary fractions are fixed. \citet{tren07} found that concentrated King models initially presents a core expansion due to the high energy injected by three- and four-body interactions, while clusters with a shallow King profile or a Plummer profile have an initial contraction. Nevertheless, both concentrated and shallow profiles tend to a common value of the core radius $r_c$. Typically, the size of the IMBH sphere of influence is proportional to the core radius of the host GC \citep{bau04a,bau04b}
\begin{equation}
r_i\propto \frac{\imbh}{M_{\rm GC}}r_c\ .
\end{equation}
As a consequence, on short timescales the extent of the cusp built up by the IMBH shall follow the the evolution of the core radius, thus expanding or shrinking according to energy injection to the half-mass radius expansion due to few-body interactions. On longer timescales, the size of the IMBH sphere of influence would tend to roughly the same value both for clusters with concentrated and shallow profiles, following the behaviour of $r_c$. As discussed in the next section, this would also affect the rate of high-velocity ejections.

\subsection{High velocity stars}
\begin{figure*}
\begin{center}
\includegraphics[width=0.9\textwidth]{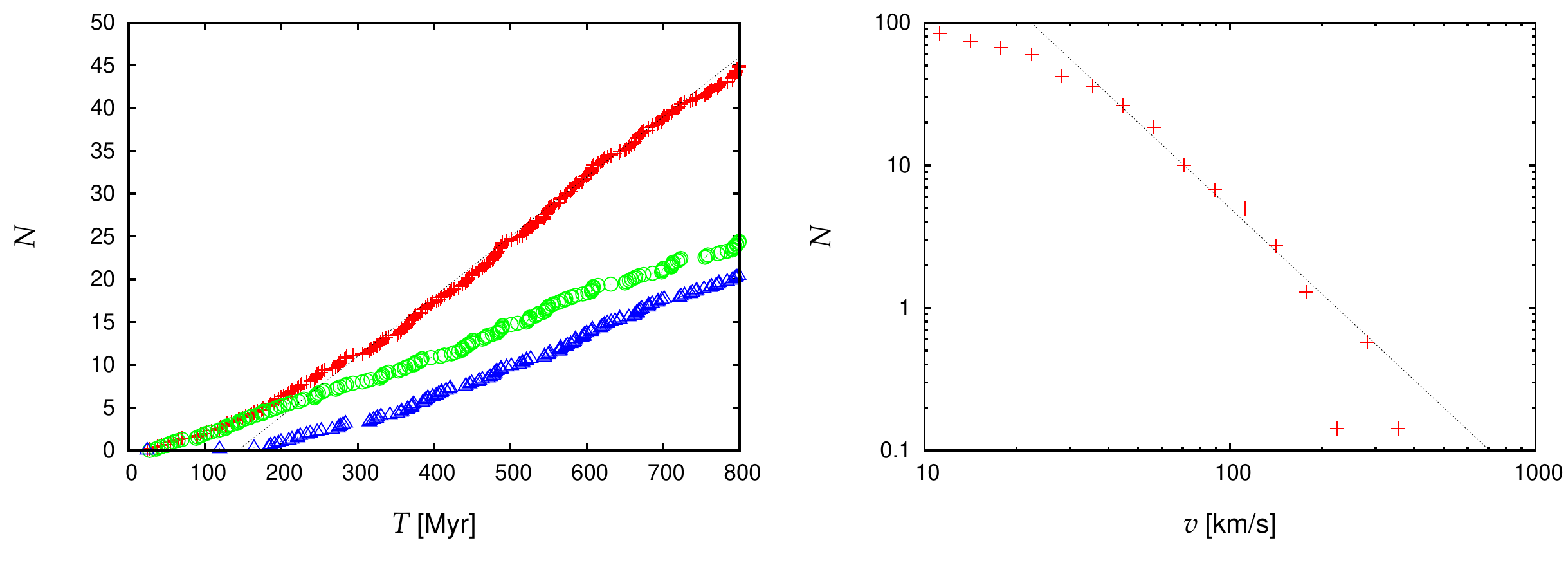}
\end{center}
\caption{\label{fig:hvs}
Properties of ejected stars from model \MODBHBIN. Left: cumulative number of stars escaping with velocities $> 50\,\kms$. Green circles correspond to main-sequence stars while blue triangles show escape rate of compact objects. Red crosses represent the total and the thin dotted line indicates escape rate of 7 stars per $100\,\myr$. Right: number counts of escaping stars up to $T = 800\,\myr$ in logarithmically equal-width bins, i.e. the displayed quantity is proportional to $\vesc\,n(\vesc)$ with $n(\vesc)$ being the distribution function of escape velocities. For $\vesc \gtrsim 30\,\kms$ the displayed quantity may be approximated with a power-law $\propto \vesc^{-2}$ (indicated by thin dotted line).}
\end{figure*}

A binary star of total mass $m=m_1+m_2$ and semi-major axis $a$ is tidally separated whenever it approaches the IMBH within the tidal radius \citep{hills88,yut03}
\begin{equation}
r_t\approx \left(\frac{\imbh}{m}\right)^{1/3} a\ .
\label{eqn:rt}
\end{equation}
The typical (average) velocity of the ejected star is \citep{brm06}
\begin{equation}
v_{ej}\approx 460 \left(\frac{a}{0.1\au}\right)^{-1/2} \left(\frac{m}{2\msun}\right)^{1/3} \left(\frac{\imbh}{10^3\msun}\right)^{1/6} \kms\ .
\label{eqn:vej}
\end{equation}
For an unequal-mass binary, due to momentum conservation the ejection speeds of the primary and secondary are
\begin{equation}
v_1=v_{ej}\left(\frac{2m_2}{m}\right)^{1/2}
\makebox[3em]{and}
v_2=v_{ej}\left(\frac{2m_1}{m}\right)^{1/2}\ ,
\label{eqn:vej2}
\end{equation}
respectively. While weakly dependent on the IMBH mass, the ejection velocity depends mostly on the binary semi-major axis: stars that were bound in tighter binaries are ejected with larger velocities. For what concerns the ejection rates, binaries can be disrupted in the full loss-cone or the empty loss-cone regime, according to the mass of the IMBH and the velocity dispersion of the surrounding environment \citep{mer13}. In the full loss-cone regime, binary stars are scattered in and out of the loss-cone along their orbital path, while any star deflected into the loss-cone is disrupted within a dynamical time, in the empty loss-cone regime. In the case of binary disruptions by an IMBH in the core of a star cluster, \citet{pfa2005} found a typical rate for the full loss-cone regime
\begin{eqnarray}
\mathcal{R} & \approx & f_b \left(\frac{a}{0.1 \au}\right) \left(\frac{n}{10^5 \pc^3}\right) \left(\frac{\imbh}{10^3 \msun}\right)^{4/3}\times\nonumber\\
& \times & \left(\frac{\sigma}{10\kms}\right)^{-1} \mathrm{Myr}^{-1}\ ,
\label{eqn:rateimbh0}
\end{eqnarray}
while, in case of empty loss-cone
\begin{equation}
\mathcal{R}\approx f_b \left(\frac{n}{10^5 \pc^3}\right)^{2} \left(\frac{\imbh}{10^3 \msun}\right)^{3} \left(\frac{\sigma}{10\kms}\right)^{-9} \mathrm{Myr}^{-1}\ .
\label{eqn:rateimbh}
\end{equation}
Here, $\sigma$ is the core velocity dispersion and $n$ the number density of the host cluster, and $f_b$ is the binary fraction. Note that while the rate depends on the binary semi-major axis in the full loss-cone regime, it is independent of in the empty loss-cone. For typical clusters and IMBH masses, the rates are $\sim 0.1$ -- $1 f_b\ \mathrm{Myr}^{-1}$. 

Our numerical integrations show that only in the model \MODBHBIN, i.e. when we include both primordial binaries and the IMBH, a considerable amount of high velocity escaping stars is being produced ($\approx 45$ stars with $\vesc \geq 50\,\kms$ in $800\,\myr$). Other models give on average less than one star with $\vesc \geq 50\,\kms$. This fact is a strong indication for that the primary mechanism that accelerates stars to high velocities is a separation of a binary star in the tidal field of the IMBH, as previously discussed.

Fig.~\ref{fig:hvs} shows distribution of the escaping stars in model {\MODBHBIN} in the time and velocity domain. Left panel displays cumulative distribution in time, i.e. it effectively provides escape rates per unit time. The escape rate is $\approx 2.5$ stars per $100\,\myr$ for $T \lesssim 200\,\myr$. Then we observe growth of the escape rate which is likely related to the buildup of the stellar cusp around the IMBH. After $T \approx 300\,\myr$ it is nearly constant with a value of $\approx 7$ stars per $100\,\myr$ (the escape rate is approximately four times larger if we push the lower limit of escaping stars to $25\,\kms$). Roughly one half of stars with $\vesc > 50\,\kms$ are ordinary stars while the second half accounts for compact objects. Considering the fact that compact objects represent only $\approx 11$ per cent of the total number of `stars' in the cluster, they are being ejected more effectively than the lighter ordinary stars. This is likely due to that the compact objects sink to the cluster centre by means of dynamical friction where the accelerator, the IMBH, resides.

Velocities of escaping stars span over more than one order of magnitude, starting from $\approx 10\,\kms$ while the highest velocity achieved in our models is slightly above $300\,\kms$. For $\vesc \gtrsim 30\,\kms$ the distribution of velocities may be approximated with a power-law, $n(\vesc) \propto \vesc^{-3}$. Combining the analytic approximations to the output of the numerical model we obtain an estimate of the rate of escaping stars per unit velocity and unit time for $\vesc \gtrsim 30\,\kms$ as
\begin{equation}
\frac{\mathrm{d} N_\mathrm{esc}}{\mathrm{d}\vesc\,\mathrm{d} t} \approx 350 \left( \frac{\vesc}{\kms} \right)^{-3} \left( \kms \right)^{-1} \myr^{-1}\;.
\label{eqn:esc}
\end{equation}

As discussed, the typical ejection velocity depends on the initial properties of the binary population, while it weakly varies with the IMBH mass. Thus, changing the IMBH would only slightly affect the typical ejection speed of the high-velocity stars. On the other hand, the choice of the binary semi-major axis distribution is fundamental in determining the typical average ejection speed of the stars ejected from the cluster \citep{perets12,subr16}, being $v_{ej} \propto a^{-1/2}$ (Eq.~\ref{eqn:vej}). In our simulations, we adopted a log-normal distribution which peaks at a semi-major axis corresponding to an orbital period of $\sim 180$ yr \citep{dm91}. An initial distribution that would favour smaller semi-major axis, as e.g. a log-uniform distribution (\"{O}pik's law), would lead to larger ejection velocities. While not affecting the ejection velocity, the mass of the IMBH sets the typical disruption rate of binaries and, as a consequence, of high-velocity escapers, as seen in Eqs.~(\ref{eqn:rateimbh0}) -- (\ref{eqn:rateimbh}). In our models, we found a total ejection rate of the order of $\sim$ one star (or compact stellar remnant) per $\sim 10\,\myr$, roughly consistent with the previous equation. Fixed the cluster properties, the rate is $\propto \imbh^{4/3}$ and $\propto \imbh^{3}$ in the full and empty loss-cone regime, respectively. A larger IMBH mass would give larger rates, thus enhancing the number of high-velocity stars ejected from the cluster. Similarly, clusters with larger central density and smaller velocity dispersion would produce a larger amount of ejected stars. This is in turn related to the characteristics of the cluster density profile, upon which a cusp is built by the IMBH, as discussed in the previous section \citep[see also][]{tren07}. We stress that the cluster binary fraction plays a crucial role in determining the rate of ejected stars, since the rate is lineraly dependent on it. Cluster harbouring an IMBH but with a small binary fraction would give a few ejections over the cluster lifetime, as found in our model \MODBH. Finally, we note that $f_b$ is not constant throughout the cluster evolution, but it decreases in time. \citet{tren07} found that typically the binary fraction decreases to $\sim 60$-$80\%$ of the initial value, both for shallow Plummer and concentrated King profile. As a consequence, the ejection rate of stars is expected to decrease with time, following the evolution of $f_b$.

\subsubsection{Population of ejected stars in host galaxy}
\label{sec:population}
\begin{figure*} 
\centering
\begin{minipage}{20.5cm}
\hspace{0.5cm}
\subfloat{\includegraphics[scale=0.5]{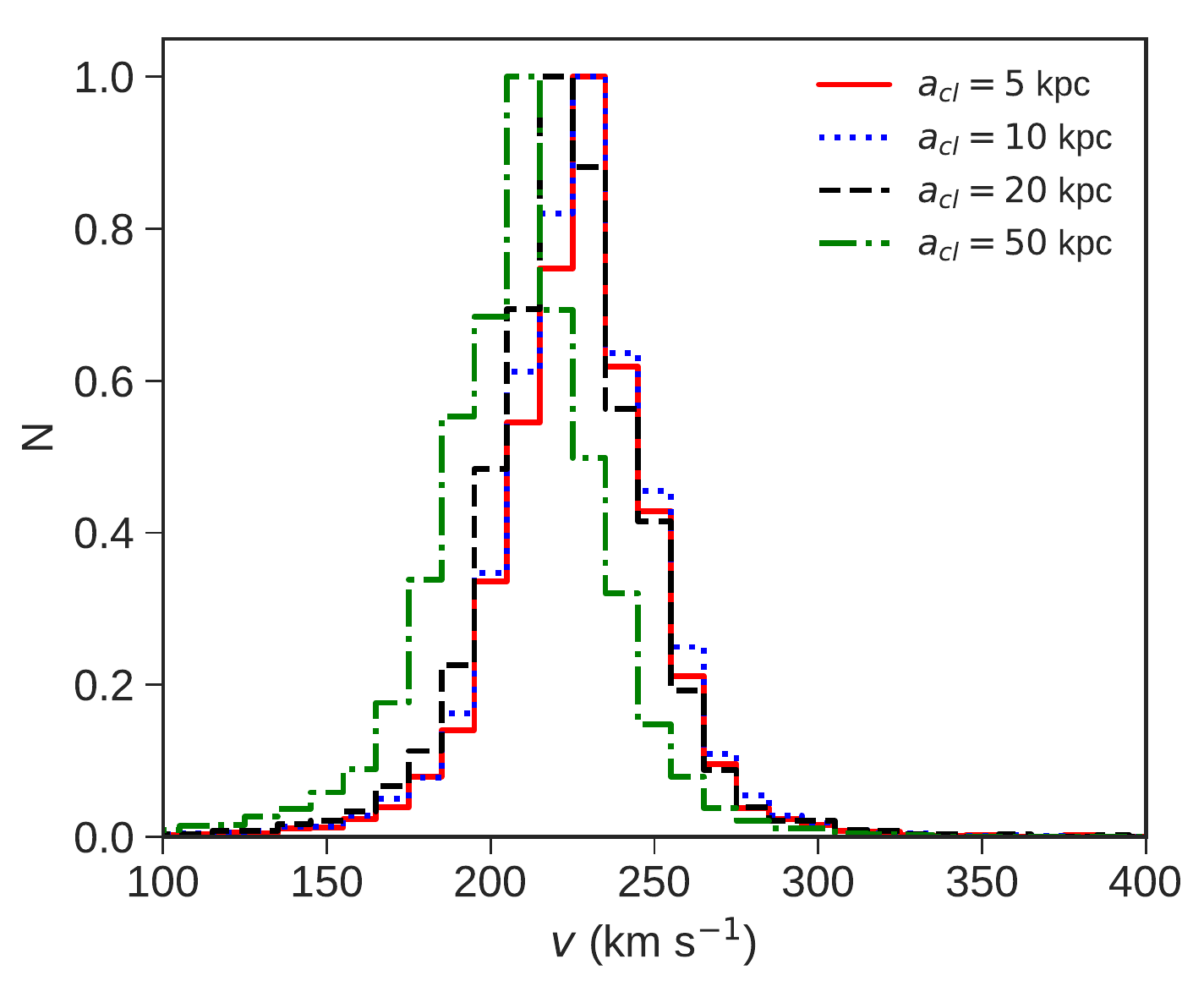}}
\hspace{1.5cm}
\subfloat{\includegraphics[scale=0.5]{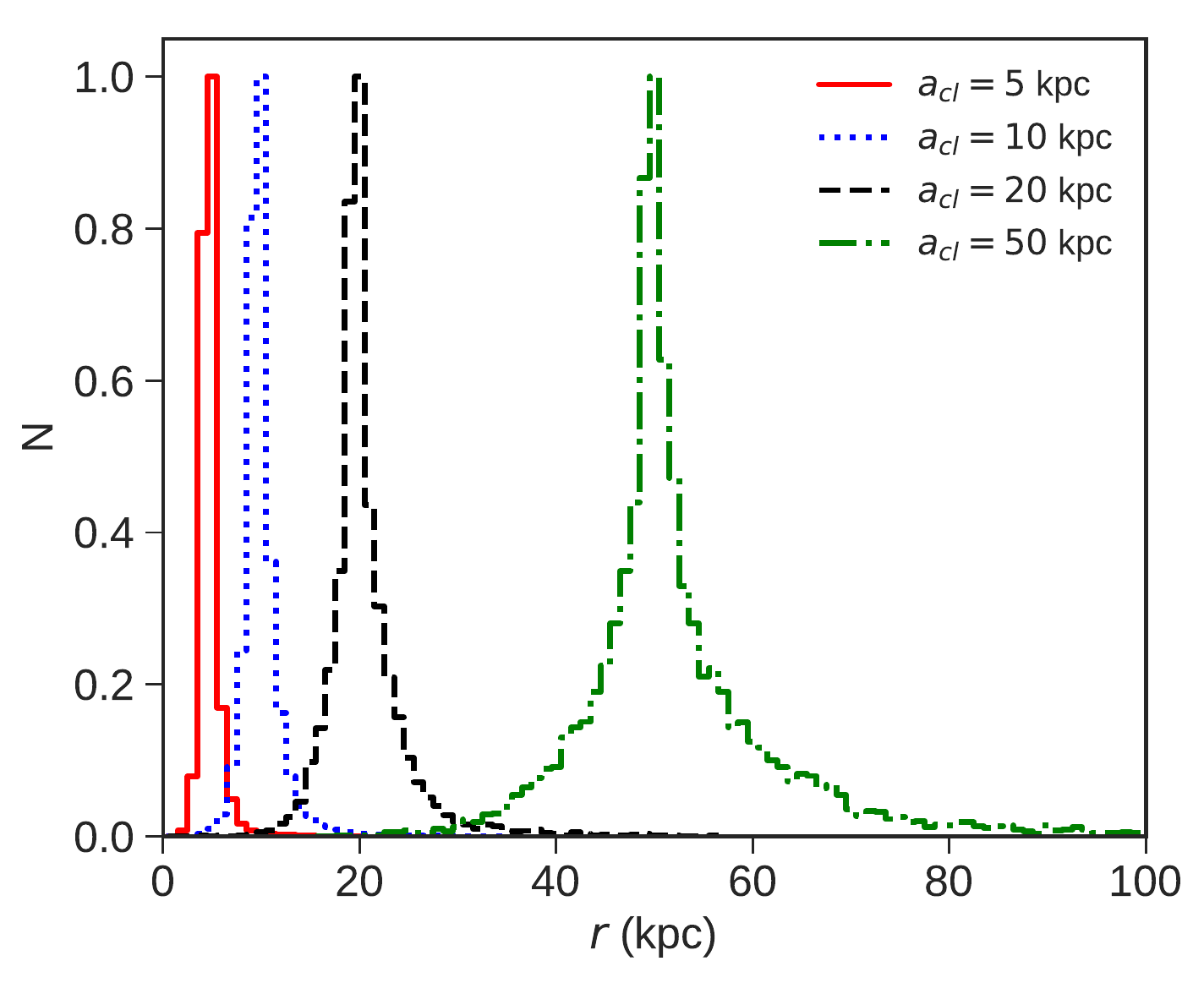}}
\end{minipage}
\begin{minipage}{20.5cm}
\hspace{0.5cm}
\subfloat{\includegraphics[scale=0.5]{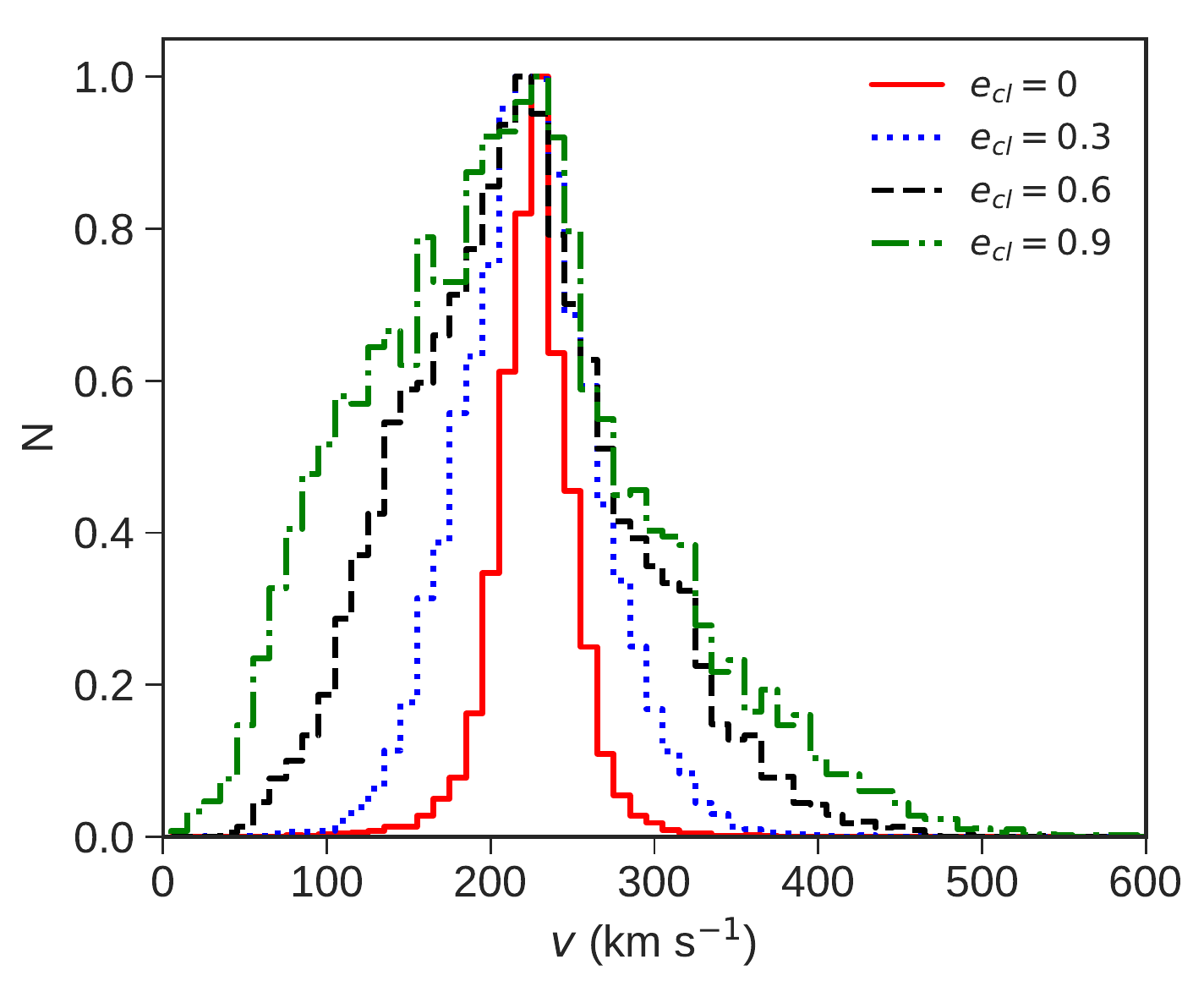}}
\hspace{1.5cm}
\subfloat{\includegraphics[scale=0.5]{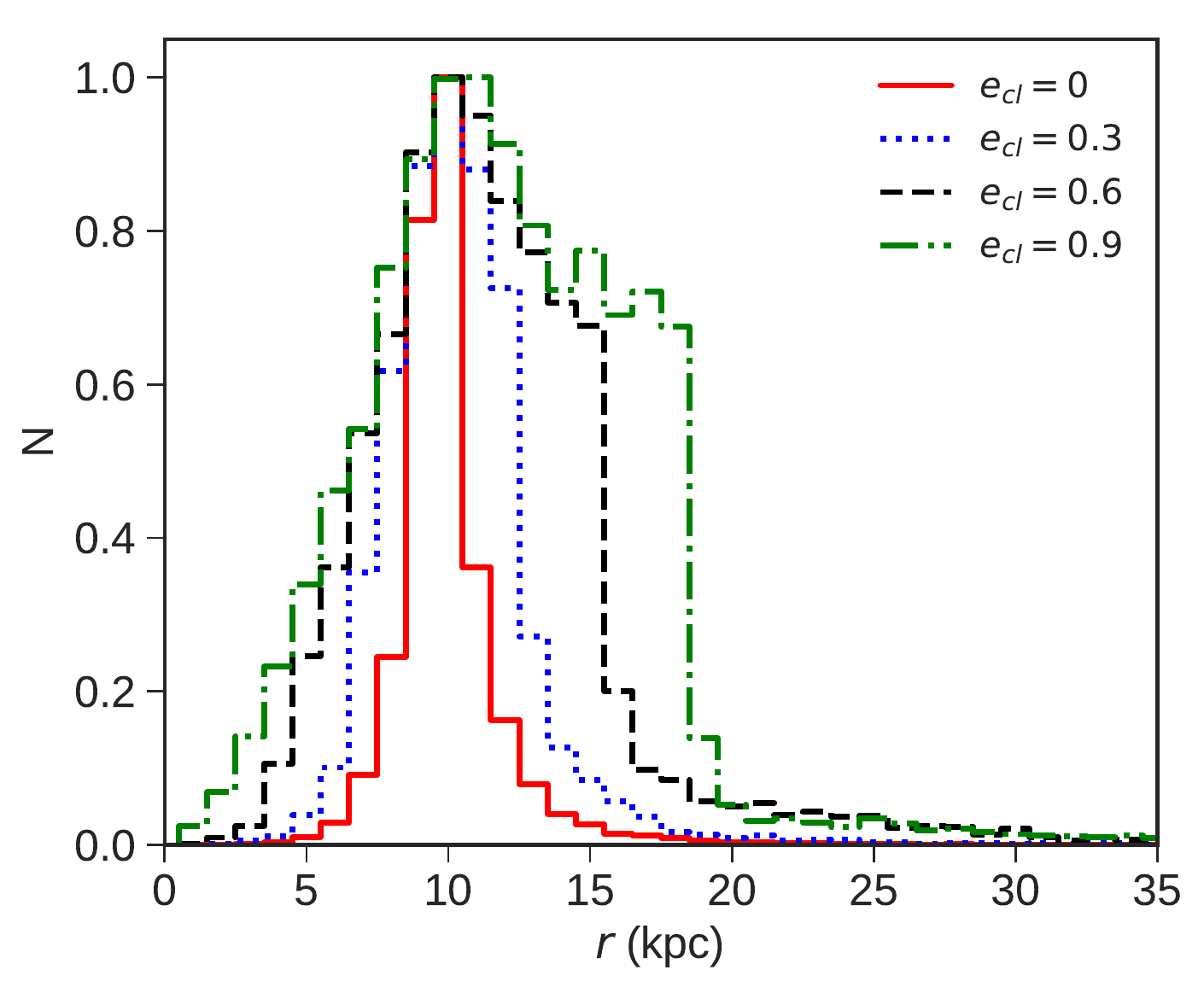}}
\end{minipage}
\begin{minipage}{20.5cm}
\hspace{0.5cm}
\subfloat{\includegraphics[scale=0.5]{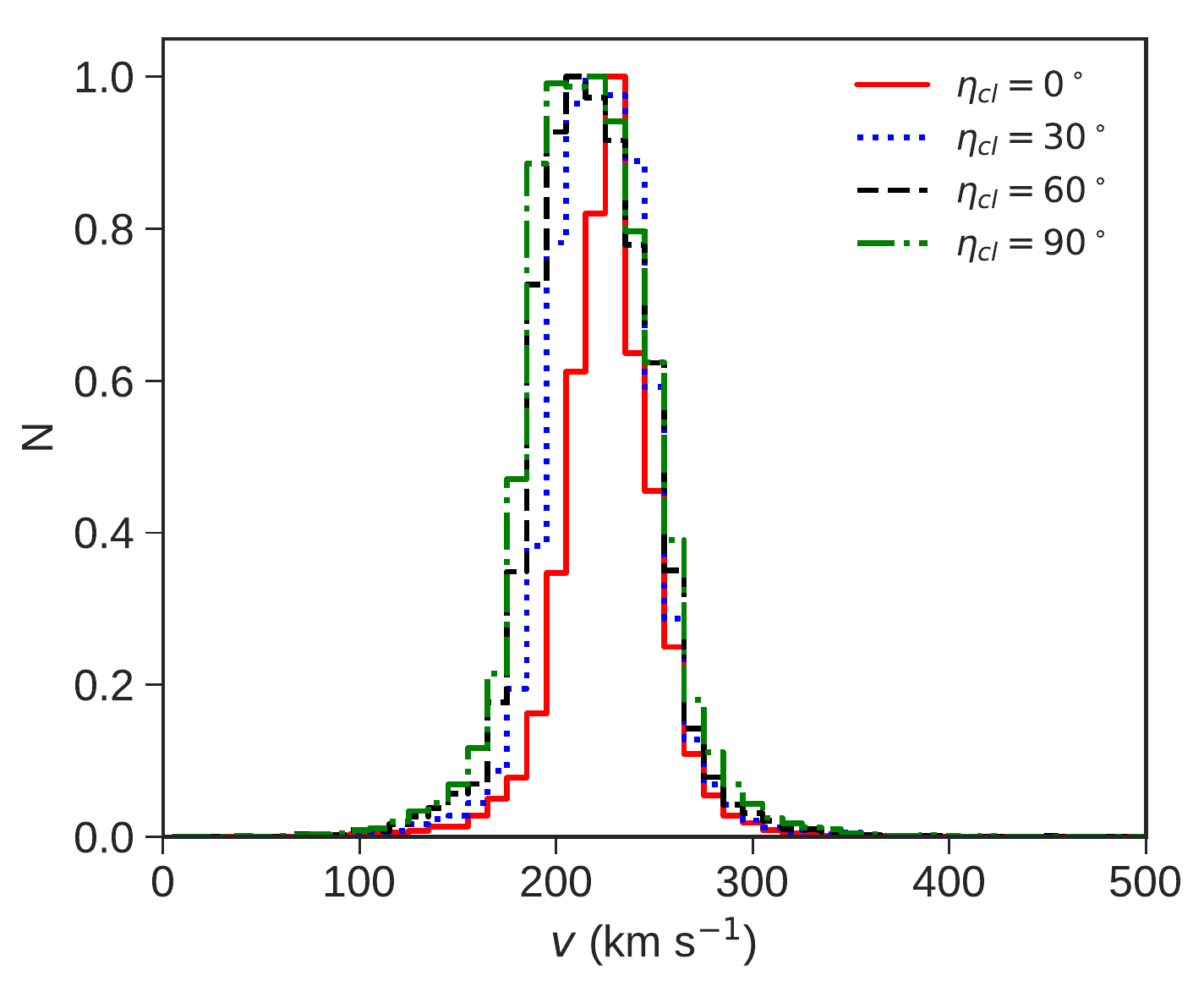}}
\hspace{1.5cm}
\subfloat{\includegraphics[scale=0.5]{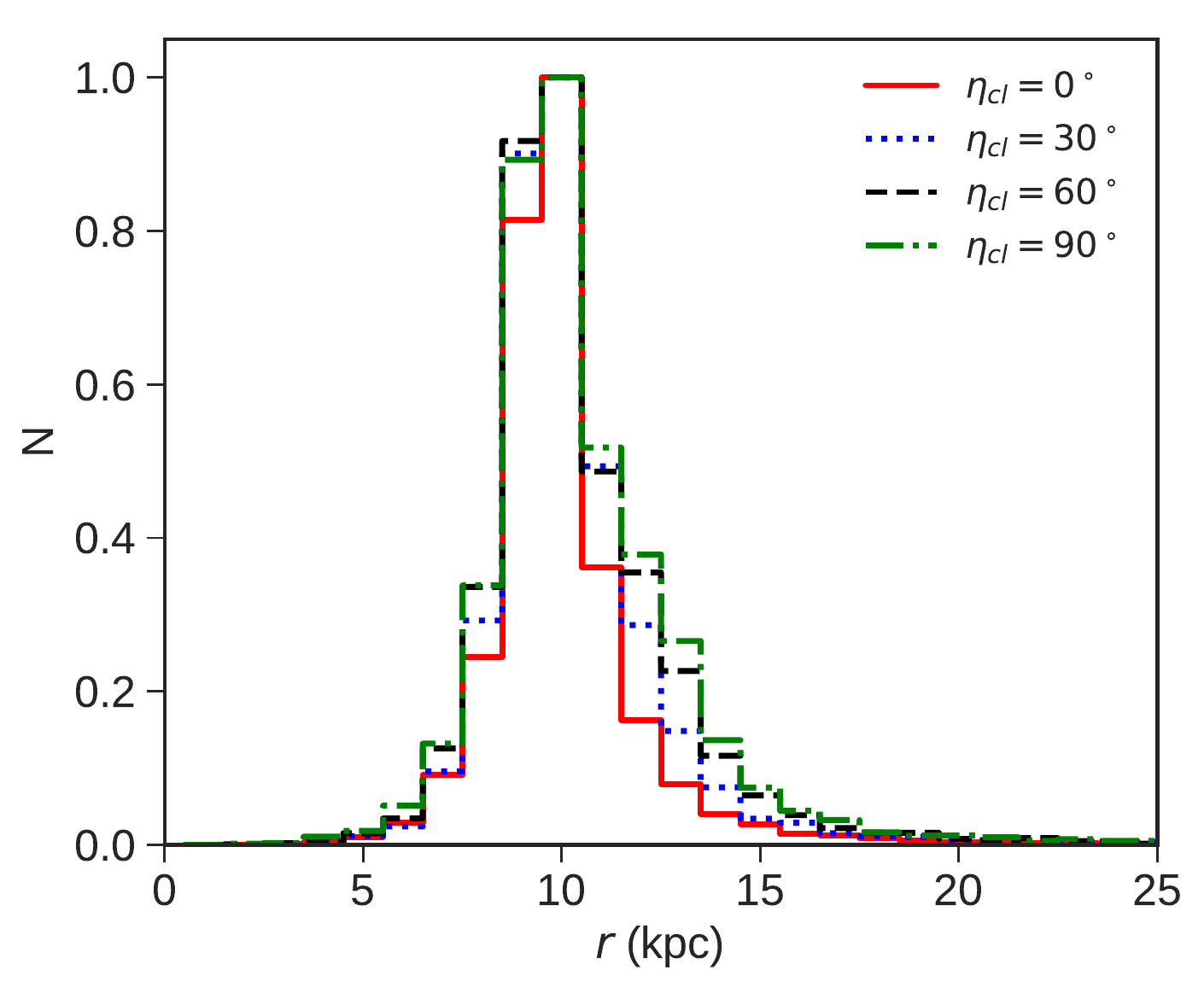}}
\end{minipage}
\caption{Arbitrarily normalised final distributions of velocity (left) and position (right) in Galactic rest frame for stars ejected from a cluster of with different orbital semi-major axis when $e_{\rm cl}=0$ and $\eta_{\rm cl}=0^\circ$ (top), eccentricities when $a_{\rm cl}=10\kpc$ and $\eta_{\rm cl}=0^\circ$ (centre) and $\eta_{\rm cl}$ when $a_{\rm cl}=10\kpc$ and $e_{\rm cl}=0$ (bottom).}
\label{fig:hvsradvel}
\end{figure*}

\begin{figure*} 
\centering
\begin{minipage}{20.5cm}
\hspace{0.5cm}
\subfloat{\includegraphics[scale=0.50]{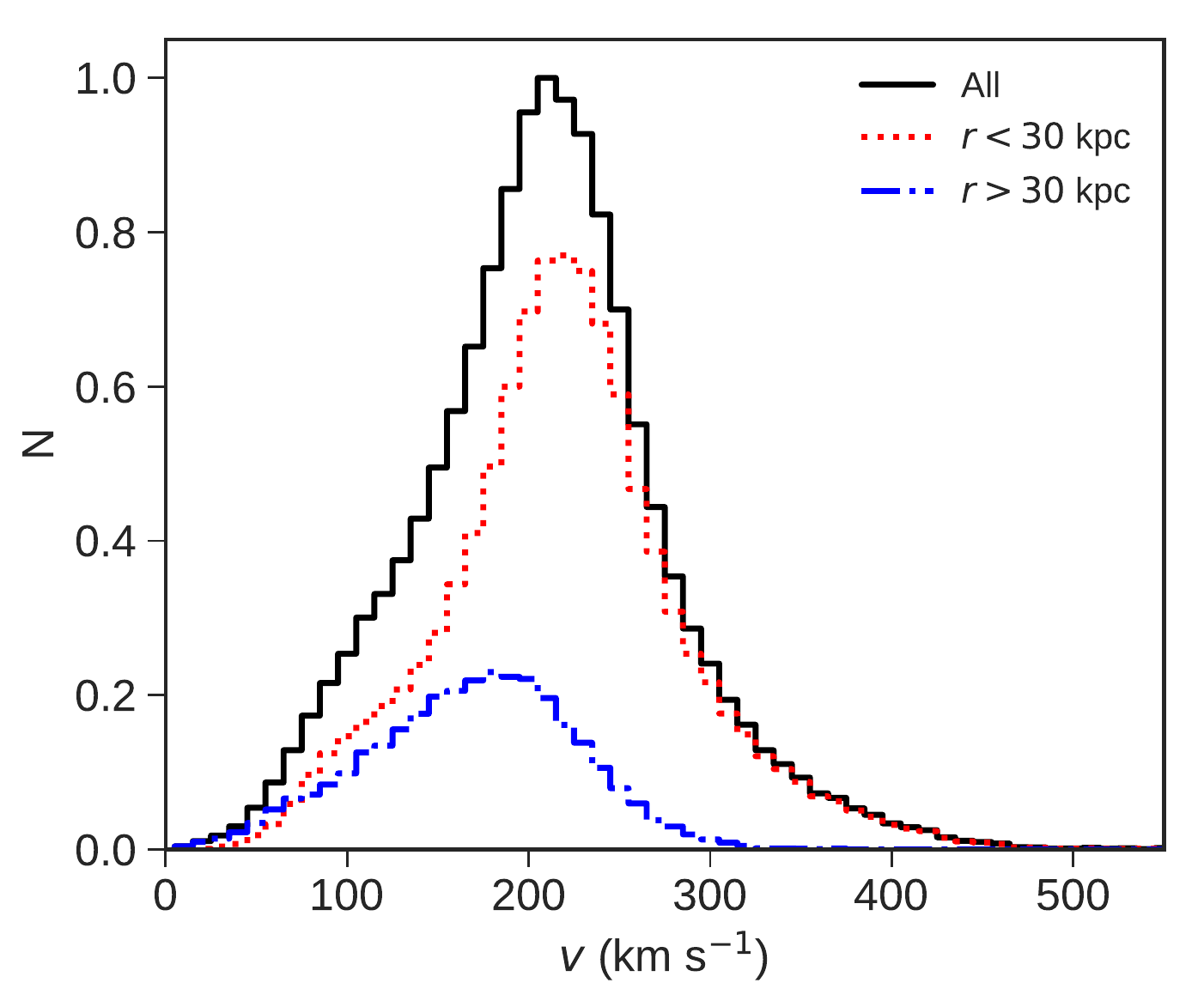}}
\hspace{1.5cm}
\subfloat{\includegraphics[scale=0.50]{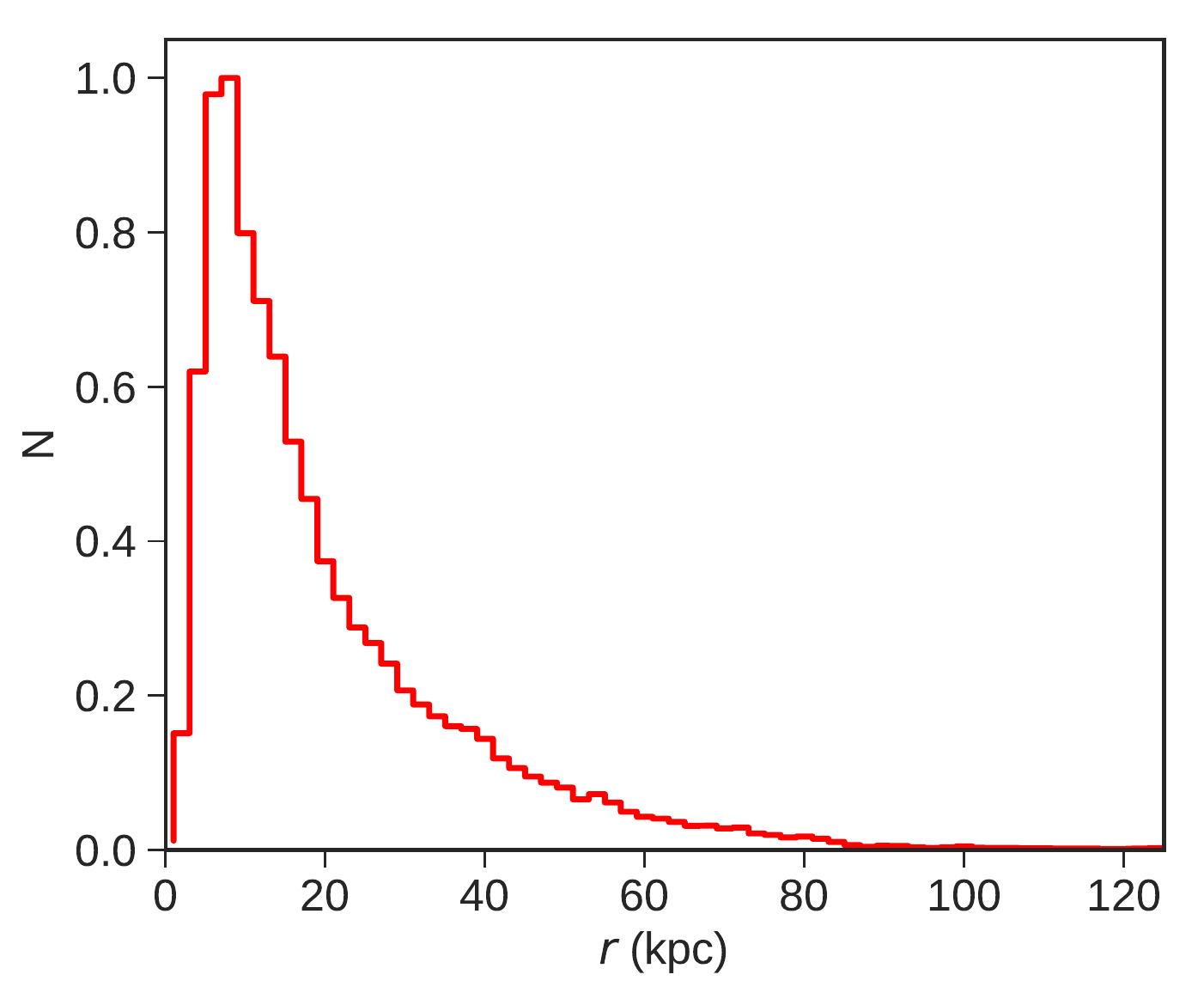}}
\end{minipage}
\caption{Final velocity (left) and spatial (right) distribution in the Galactic rest frame for stars ejected by 100 clusters in a Milky-Way like galaxy. Cluster semi-major axis are drawn from a log-uniform distribution ($a_{min}=5\kpc \le a_{cl}\le a_{max}=50\kpc$), eccentricities from a thermal distribution and inclinations from an isotropic distribution (uniform in $\cos \eta_{\rm cl}$).} 
\label{fig:hvsradvelall}
\end{figure*}

Although we have studied a limited number of clusters, we use the inferred rate (Eq.~\ref{eqn:esc}) as a proxy to estimate the statistical properties of the population of stars ejected from GCs in a Milky Way-like galaxy. Even though the most important features of ejected stars are captured by Eqs.~(\ref{eqn:vej}) -- (\ref{eqn:rateimbh}), further studies that include a wider span of initial conditions, importantly the cluster and IMBH size and the distribution function of binary semi-major axis, are necessary to study in detail how different clusters populate a Milky Way-like (or another) galaxy with high-velocity stars.

First, we investigate ejection of high-velocity stars from a single cluster hosting an IMBH that moves in the galactic gravitational field. We characterize the cluster orbit by means of three parameters \citep{frgual18b}: (i) semi-major axis $a_{\rm GC}$, (ii) eccentricity $e_{\rm GC}$ and (iii) relative inclination of the cluster orbital plane and of the Galactic disc $\eta_{\rm cl}$ ($\eta_{\rm cl}=0^\circ$ corresponds to an orbital plane coinciding with the Galactic disk).

The Milky Way-like galaxy potential is described with a 4-component model $\Phi(r)=\Phi_{\rm BH}+\Phi_{\rm b}(r)+\Phi_{\rm d}(r)+\Phi_{\rm h}(r)$ \citep{ken14,frl17}, where:
\begin{itemize}
\item $\Phi_{\rm BH}$ is the contribution of the central SMBH,
\begin{equation}
\Phi_{\rm BH}(r)=-\frac{GM_{\rm BH}}{r},
\end{equation}
with mass $M_{\rm BH}=4 \times 10^6 \ {\rm M}_{\odot}$;
\item $\Phi_{\rm b}$ is the contribution of the spherical bulge,
\begin{equation}
\Phi_{\rm b}(r)=-\frac{GM_{\rm bul}}{r+a},
\end{equation}
with mass $M_{\rm b}=3.76\times 10^9 \ {\rm M}_{\odot}$ and scale radius $a=0.10\kpc$;
\item $\Phi_{\rm d}$ accounts for the axisymmetric disc,
\begin{equation}
\Phi_{\rm disk}(R,z)=-\frac{GM_{\rm disk}}{\sqrt(R^2+(b+\sqrt{c^2+z^2})^2)},
\end{equation}
with mass $M_{\rm disk}=5.36\times 10^{10} \ {\rm M}_{\odot}$, length scale $b=2.75\kpc$ and scale height $c=0.30\kpc$;
\item $\Phi_{\rm halo}$ is the contribution of the dark matter halo
\begin{equation}
\Phi_{\rm halo}(r)=-\frac{GM_{\rm DM}\ln(1+r/r_s)}{r}.
\end{equation}
with $M_{\rm DM}=10^{12} \ {\rm M}_{\odot}$ and length scale $r_{\rm s}=20\kpc$.
\end{itemize}
The potential parameters are set so that the Galactic circular velocity is $235 \ {\rm km \, s}^{-1}$ at the Sun's distance ($8.15\kpc$).

To generate mock populations of ejected stars, we follow \citet{ken14} prescriptions. Moreover, we assume that the IMBH ejects stars at a constant rate along the cluster orbit. When we eject a star, we randomly draw an ejection time $t_{\rm ej}$ and an observation time $t_{\rm obs}$ between zero and the star's main-sequence lifetime. In this approach, we both account for the fact that the star interacts with the IMBH and is observed before it evolves off the main-sequence. If $t_{\rm ej}<t_{\rm obs}$, we assign the star an ejection velocity $v_{\rm esc}$ sampled from equation~(\ref{eqn:esc}). Since we are interested in how the ejected stars populate the host galaxy, we combine the star ejection velocity with the cluster orbital velocity at the moment of ejection. We then integrate the star orbit across the host galaxy up to a maximum time $\mathcal{T}=t_{\rm obs}-t_{\rm ej}$. If at any time the star passes the virial radius (250 kpc), we consider the star ejected from the galaxy and we remove it from our calculation.

Fig.~\ref{fig:hvsradvel} illustrates the final distribution of velocity (left) and position (right) in Galactic rest frame for stars ejected from a cluster with different orbital semi-major axis when $e_{\rm cl}=0$ and $\eta_{\rm cl}=0^\circ$ (top), eccentricities when $a_{\rm cl}=10$ kpc and $\eta_{\rm cl}=0^\circ$ (centre) and $\eta_{\rm cl}$ when $a_{\rm cl}=10$ kpc and $e_{\rm cl}=0$ (bottom). The relative inclination $\eta_{\rm cl}$ between the cluster orbit and the Galactic disc does not play an important role in determining the final shape of the velocity and position distribution. On the other hand, the cluster orbital semi-major axis affects the final spatial distribution, which is peaked around the cluster semi-major axis, but not the final velocity distribution that results peaked near the cluster orbital velocity. Finally, the cluster eccentricity affects both of the distributions in velocity and position: the larger the eccentricity the larger the broadening around the peaks determined by the cluster orbital semi-major axis.

\begin{figure*}
\begin{center}
\includegraphics[width=0.98\textwidth]{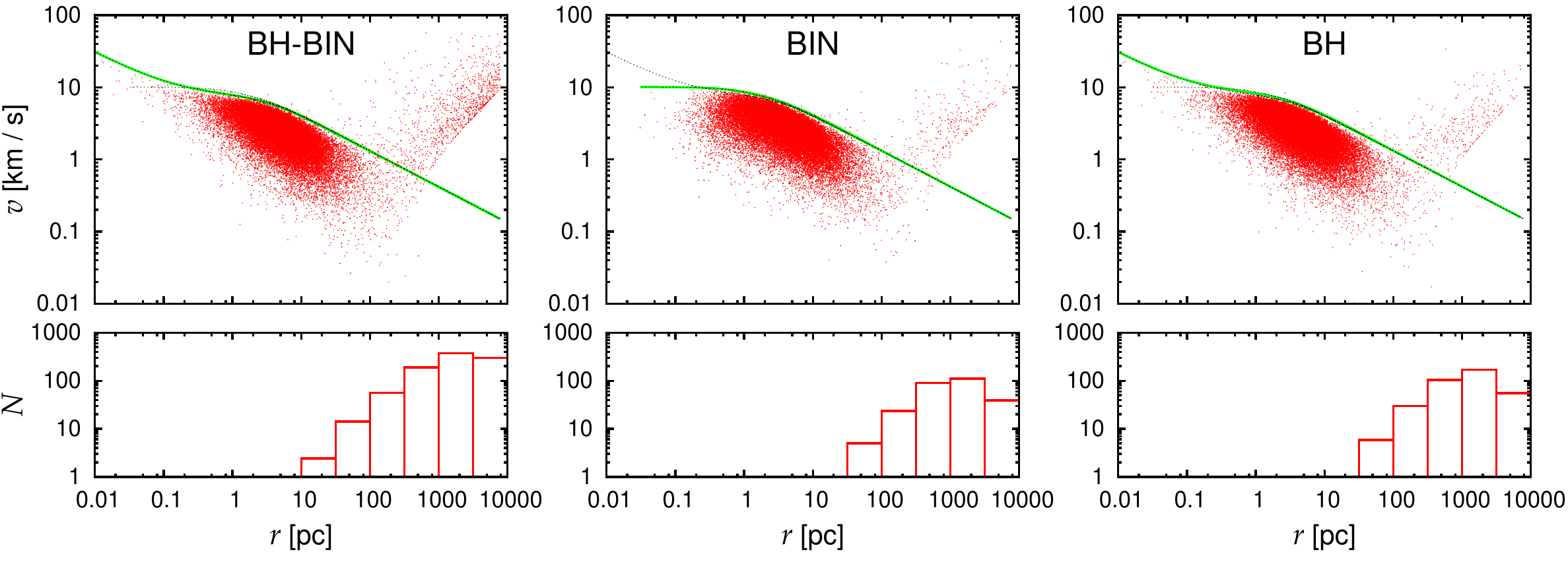}
\end{center}
\caption{\label{fig:rv}
Top panels: Velocity vs. radial distance from the cluster centre at $T = 800$ Myr for all three models of star clusters (one randomly selected run per model). Thick green line indicates escape velocity at given radius determined by the method of H\'enon~(1971); thin dotted lines in each panel show escape velocity limits for all other models for comparison (the lines for models {\MODBH} and {\MODBHBIN} lie on the top of each other). Bottom: number counts of stars with velocity $v>1.5\vesc$ in logarithmically equal-width radial bins.}
\end{figure*}

\begin{figure}
\begin{center}
\includegraphics[width=0.99\columnwidth]{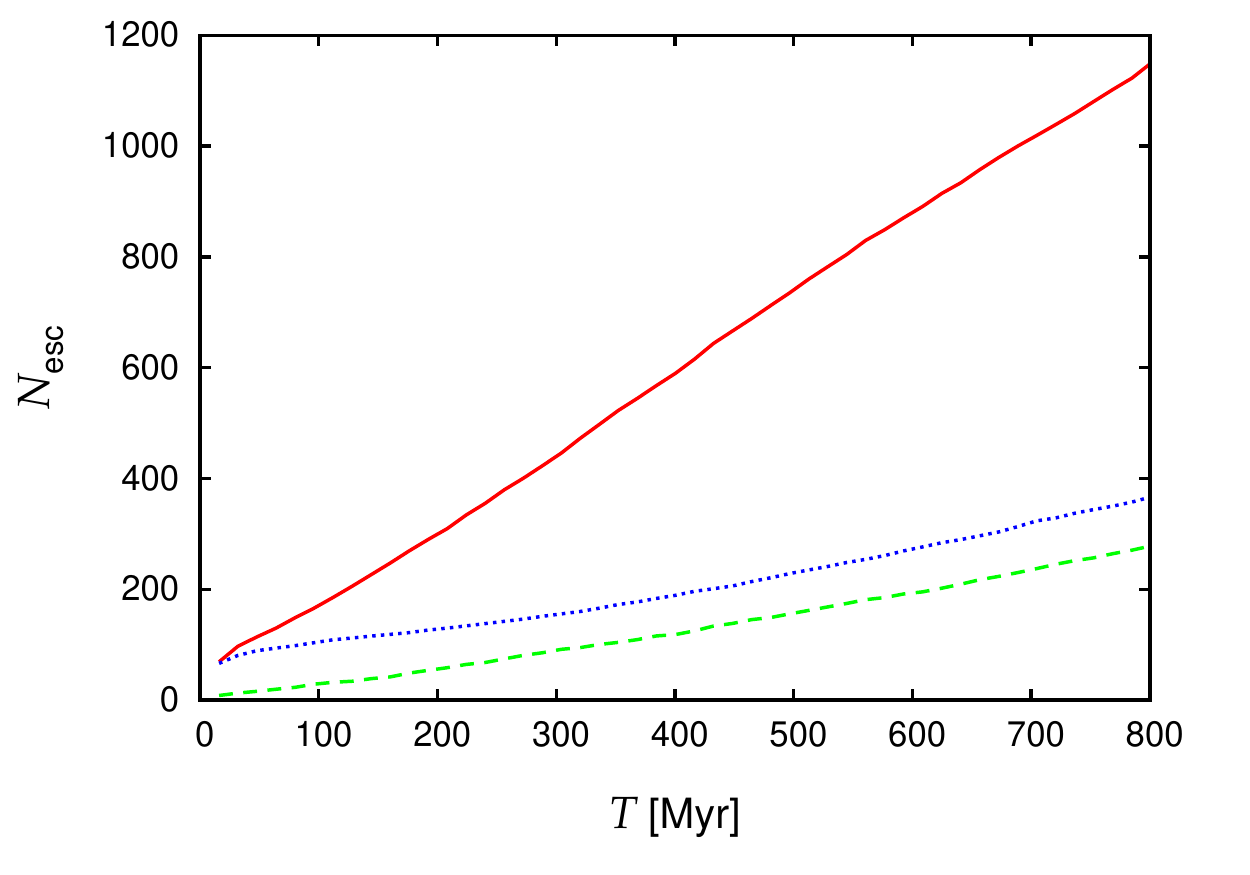}
\end{center}
\caption{\label{fig:evaporate}
Temporal evolution of the number of escaping (unbound; $v>1.5\vesc$) stars for models {\MODBHBIN} (solid red line), {\MODBH} (dotted blue) and {\MODBIN} (dashed green).}
\end{figure}

To understand how the stars ejected from a distribution of clusters populate the host galaxy, we use the previous scheme by considering a population of 100 clusters with different semi-major axis, eccentricities and inclinations. We sample cluster semi-major axis from a log-uniform distribution (between $a_{\rm min}=5\kpc$ and $a_{\rm max}=50\kpc$), eccentricities from a thermal distribution and inclinations from an isotropic distribution (uniform in $\cos \eta_{\rm cl}$). Fig.~\ref{fig:hvsradvelall} illustrates the final velocity (left) and spatial (right) distribution of the ejected stars. Most of the ejected stars are located within $30$ kpc and have peak velocity $\approx 250\kms$, with a tail extending up to $\approx 500\kms$. Our model does not produce a large population of stars with extreme velocities (as a consequence of the assumed initial distribution for binary periods). Hence, we can not account for most of the unbound HVS population observed in the Milky Way \citep{brown14,brw15}. However, we note that the main feature of our model is to produce fast moving stars that do not point back towards the Galactic Centre, so stars have a significant non-radial component of motion (in Galactic coordinates), as in the case of runaways and hyperunaways produced in the Galactic disc \citep{silva11,pallad14,zhong14,vickers15}. Interestingly, recent analysis of \textit{Gaia}\footnote{http://sci.esa.int/gaia/} data showed that only a few of the confirmed and candidate HVSs have orbits that can be traced back to the Galactic Centre \citep{boub18,brown2018,march18}.

Studying RS and HVS kinematic and spectroscopic data can help constraining their origin and possibly the presence of an IMBH. Spectroscopic data can be used to get radial velocities, while tangential velocities can be obtained from proper motion measurements. The combination of radial and tangential velocity, along with the position in the sky of the stars, gives the full 6-D phase space information to study the stars' orbits. \citet{hoo01} used milli-arcsecond accuracy astrometry from \textit{Hipparcos} and from radio observations of the orbits of $56$ RSs and nine compact objects with distances $\lesssim 700$ pc, to identify their parent stellar group. \citet{hebe08} studied the mass, evolutionary lifetime and kinematics of HD 271791 by using proper motion measurements from a collection of catalogues. They found that the likely birthplace is the outer Galactic disc, while the Galactic Centre is ruled out. Recently, \citet{lenno18} and \citet{renz18} used \textit{Gaia} data to show that the dynamics of the very-massive runaways VFTS 16 and VFTS682 are consistent with them having been ejected from the young massive cluster R136. Finally, also \citet{hatt18} used \textit{Gaia} data to study the origin of  hyper-runaway subgiant LAMOST-HVS1, and found that it was likely ejected from near the Norma spiral arm dynamically as a consequence of a few-body encounter or a Hills ejection by an IMBH. Thanks to the high precision of \textit{Gaia} proper motion similar studies can be performed for known and candidate high-velocity stars, that might have been ejected through the mechanism discussed in this work, thus disclosing new IMBH candidates.

\subsection{Low velocity escapers}

While the stars ejected from the star clusters at high velocities may have a specific appeal, their number counts are relatively small and, therefore, it may be difficult to make direct relations of high velocity stars and star clusters in observational data. In this situation, their low velocity counterparts may serve as an important diagnostic tool as they are more numerous and are found closer to their parent clusters. Our simulations indicate (see Fig.~\ref{fig:rv}) that star clusters hosting an IMBH and a considerable population of primordial binaries produce larger number of stars escaping from the cluster with velocities $1\,\kms \lesssim \vesc \lesssim 30\,\kms$ than their counterparts without an IMBH or a primordial binary population. In Fig.~\ref{fig:evaporate} we plot the temporal evolution of the number of stars with velocities greater than $1.5\vesc$ as a rough measure of the cluster evaporation rate. Here, $\vesc$ is determined by a method developed by \cite{henon71a} for Monte Carlo simulations of spherically symmetric star clusters: At the radius $r_i \leq r \leq r_{i+1}$, with $r_i$ being ordered radial distances of stars from the cluster centre, the escape velocity is determined from the mean potential,
\begin{equation}
2\vesc^2(r) = -V(r) = \frac{G}{r} \sum_{j=1}^{i}m_j + \sum_{j=i+1}^{N}\frac{Gm_j}{r_j}\;,
\end{equation}
where $G$ stands for the gravitational constant and $m_i$ are the masses of the individual stars. Fig.~\ref{fig:rv} indicates that this method approximates the mean potential of not exactly spherical $N$-body system quite well, although it cannot determine exactly whether a particular star is bound to the star cluster or not. Hence, the factor of $1.5$ used for the detection of evaporating stars. Yet another source of possible mis-detections are binary systems -- for those being regularised in the \code{NBODY6} code, we consider their centre of mass velocities. For weakly bound binaries, however, their full orbital velocities are used, which may be higher than $1.5\vesc$ although the binary system itself may still be bound to the star cluster.

Assuming that the primordial binary population is present in real star clusters, more or less rich tidal tails of GCs may serve as an indicator for presence of the IMBH.

\subsection{Direct interactions with the IMBH}
\begin{figure*}
\begin{center}
\includegraphics[width=0.9\textwidth]{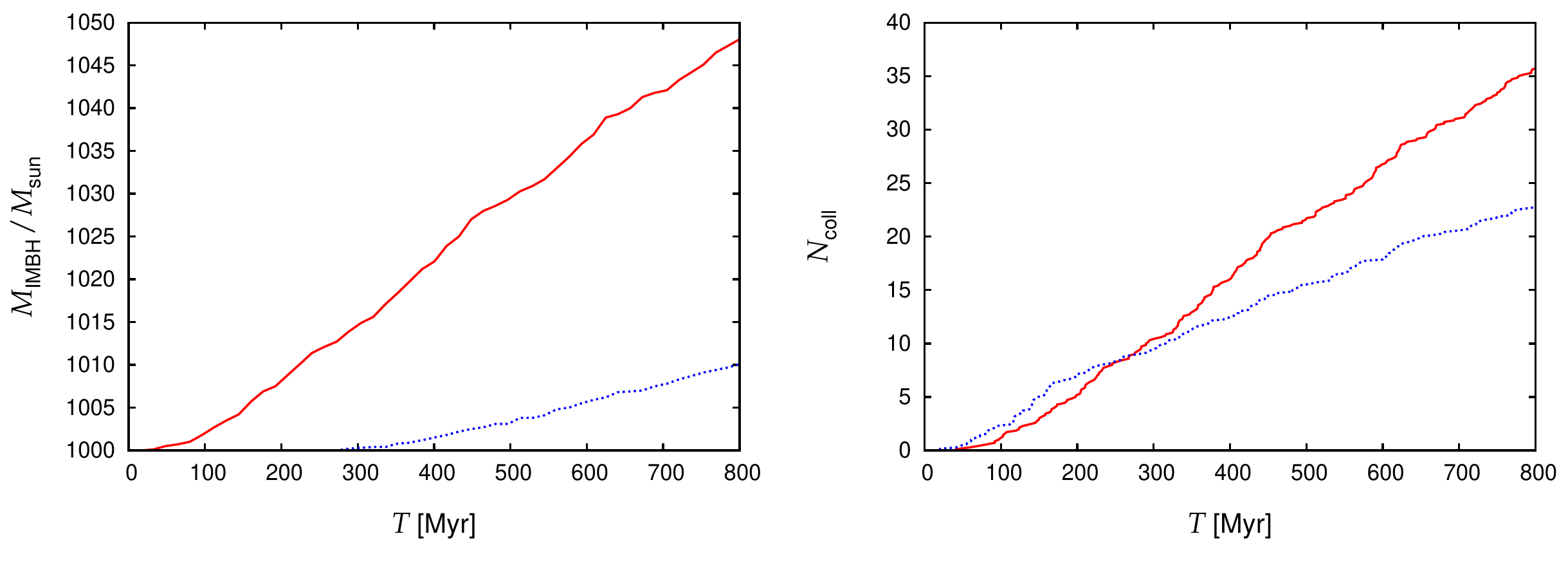}
\end{center}
\caption{\label{fig:imbh}
Left: Temporal evolution of mass of the IMBH for models {\MODBHBIN} (solid red line) and {\MODBH} (dotted blue) under assumption that all close encounters of stars and compact objects contribute to the mass of the IMBH. Right: Cumulative distribution of close encounters of stars (dotted blue line) and compact objects (solid red) with the IMBH in model {\MODBHBIN}.}
\end{figure*}

In our numerical setup, we treat `stellar' collisions in a quite simplified way, considering only two different classes: collisions that involve the IMBH and collisions that do not. For the collisions that do not involve the IMBH, the physical radii of the stellar mass objects are initially set up as $R_\ast = 0.5\rsun (\mstar / \msun)$. If the radial separation of any of two stellar mass objects becomes less than the sum of their radii, they are merged into one object whose mass is the sum of masses of the merging objects (i.e. without any mass loss). Up to the factor $0.5$, the relation used here corresponds to a classical formula for radii of low mass stars \citep[see e.g.][]{lang74}. The factor $0.5$ in the present prescription decreases the distance that leads to a merger, and is to ensure that the encountering stars really merge instead of merely having a close passage. Note that while this prescription may be accepted as a raw approximation for collisions of main sequence stars, it is definitely not adequate for stellar mass compact objects (WDs, NSs, BHs). Still, this is not a big issue, as even though there are collisions among compact objects occurring in our calculations, they are not frequent enough to considerably alter the mass spectrum and, consequently, the overall results.

For the collisions that do involve the IMBH, the criterion that distinguishes a merger from a fly-by is that the stellar mass object (i.e. a star or a stellar remnant) comes closer to the IMBH than the tidal radius of a main-sequence star of that mass. In that case, similar to the case of encounters without the IMBH, it is swallowed completely by the IMBH, regardless of its type and without any mass loss. Again, this prescription may be considered as very raw, but still plausible for main sequence stars \citep[see, e.g.,][for more rigorous description of tidal disruption events]{guillochon13}. However, it may not be adequate for compact objects, as the real tidal disruption radii for both WDs and NSs are orders of magnitudes smaller than the tidal radii of main-sequence stars, which is considered as the decisive criterion here. For NSs and stellar mass BHs, one would also have to consider gravitational radiation which would help to drag them towards the IMBH, but post-Newtonian dynamics is switched off in our calculations. Despite this caveat, we plot growth of the IMBH through mergers with stars and compact objects for models {\MODBH} and {\MODBHBIN} in left panel of Fig.~\ref{fig:imbh}. What may be interesting and worth studying with more elaborated models is the fact that the rate of growth of the IMBH roughly correlates with ejection rate of high velocity stars. Not only it saturates at a constant rate which at least by the order of magnitude corresponds to the ejections in model {\MODBHBIN}, but the correlation also goes across the models in that model {\MODBHBIN} not only produces the most high velocity stars, but also exhibits the considerable growth of the IMBH, while essentially no growth is found for model {\MODBH}.

Thus, in our simple setup, collision rates of stars / compact objects with the IMBH are rather straightforwardly related to the growth rate of the IMBH. The right panel of Fig.~\ref{fig:imbh} shows the cumulative distribution of close encounters with the IMBH for model {\MODBHBIN} distinguished for stars and compact objects. Trends found in this plot qualitatively agree with those found in the left panel of Fig.~\ref{fig:hvs}, i.e., the ejection rate as well as the rate of close encounters is initially smaller for the compact objects than for the main-sequence stars, while their relation is opposite at the end of the integrations. In the case of star--IMBH interactions, the close encounter rate may be interpreted as a rate of potentially observable tidal disruption events. The interpretation is much less straightforward for compact objects as those would require some mechanism that brings them even closer to the IMBH (e.g., two-body relaxation and/or emission of gravitational waves) in order for them to merge with the IMBH. For example, white dwarfs that dominate population of compact stellar-mass objects in real star clusters have tidal radii approximately by a factor of $100$ smaller than what was considered in our simulations.

However, all models that we consider imply that the IMBH grows only slowly through stellar dynamical processes in the environment that we simulate (a moderately old and moderately massive star cluster). This means that an IMBH is unlikely to grow from a small seed through stellar dynamics in this setting. This does not exclude that other conditions may be more favourable for the rapid growth of an IMBH, like for instance in very young clusters that still contain massive stars. Massive stars are more extended and therefore have a larger cross section for collisions. \cite{por02} found indeed in their $N$-body simulations of young star clusters that run-away mergers of massive stars can lead to objects that qualify as IMBH-progenitors. It may also be possible that a substantial growth of the IMBH can continue until later times in much more dense and massive clusters, like massive GCs and UCDs.

\section{Discussions and Conclusions}

In this paper, we have discussed the dynamical interaction of a GC of moderate mass and density with an intermediate-mass black hole. The main advancement of this study in comparison to earlier studies is that we include simultaneously to the IMBH a large population of binaries into our models for the GC, which makes the computations very demanding. Our main results are:
\begin{itemize}
\item 
All models presented in this paper started with identical density and velocity dispersion profiles given by the Plummer distribution. Clusters hosting the IMBH evolve towards power-law density profile with slope $\approx -7/4$ which is practically identical for both cases (i.e. regardless of the stellar multiplicity). At the end of our integrations ($T \approx 0.8\,\gyr$) the profiles of cluster with and without IMBH considerably differ for $r \lesssim \rh$; most prominent is the difference below $0.1\,\rh$.
\item 
In our calculations, close interactions with the IMBH led to direct increase of its mass (and removal of the interacting star from the integration). This approach definitely overestimates the growth rate of the IMBH, i.e. our results can only serve as an upper estimate for evolution of mass of the IMBH. Considering that in the most prominent case, model {\MODBHBIN} with the large binary fraction, the IMBH grew from $1000\,\msun$ to $\approx 1050\,\msun$ over $0.8\,\gyr$. This indicates that at least for lightweight GCs, the collisions of stars with the IMBH are not able to increase its mass by a factor of two on cosmological time-scales.
\item 
The most prominent feature of the model which included both the IMBH and large fraction of stellar binaries is ejection of high-velocity stars. As the ejection rate was about two orders of magnitude smaller for models either with the IMBH but no primordial binaries or with primordial binaries but no IMBH, it is clear that the acceleration is due to interaction of binary stars with the IMBH through the \cite{hills88} mechanism. The maximum velocity we observed in our model was slightly above $300\,\kms$, but this was rather exceptional case. The distribution of velocities of stars escaping from the cluster could be approximated by a power-law with a slope of $\approx -3$ (see equation~\ref{eqn:esc}) for $\vesc \gtrsim 30\,\kms$. Total ejection rate in this velocity interval is of the order of one star (or compact stellar remnant) per $10\,\myr$. This rate, together with the fact that these stars pass a distance $>300\,\pc$ over $10\,\myr$, indicates that there is quite a low probability of finding the high velocity escapers in the vicinity of parent cluster, i.e. serving possibly as indicators of the presence of the IMBH.
\item 
The high velocity stars ejected from star clusters hosting an IMBH would contribute to the overall distribution of high-velocity stars in the host galaxy. The composite distribution of high-velocity stars from a population of cluster orbiting the host galaxy (semi-major axis from a log-uniform distribution between $a_{min}=5$ kpc and $a_{max}=50$ kpc and thermal eccentricities) generate stars that are mostly within $\sim 30$ kpc with peak velocity of $\sim 250\,\kms$ and a tail extending up to $\sim 500\kms$. Since our model does not produce a large population of stars with extreme velocities, we can not account for most of the unbound HVS population observed in our Galaxy, but our mechanism can produce runaway and hyperunaway stars.
\item 
The models of star clusters that differ just by presence of the IMBH ({\MODBIN} vs {\MODBHBIN}) considerably differ by amount of stars being accelerated above the escape velocity (see Figs.~\ref{fig:rv} and \ref{fig:evaporate}). Beside the high velocity escapers which are relatively rare, model with the IMBH ({\MODBHBIN}) produces a relatively numerous population of low velocity escapers with velocities $\gtrsim 1\,\kms$. These form a diluted extended halo of the cluster which may be potentially observable for sufficiently isolated star clusters (such that cluster stripping by the galactic tidal field does not overlay the evaporation caused by the cluster internal dynamics).
\end{itemize}

Our current results are restricted to star clusters of moderate mass and density and also cover only a fraction of their life-time. This limitation stems from a large numerical complexity of our models and moderate computational resources available (one integration of the model {\MODBHBIN} took approximately half a year of computer time on a 40 CPU core sytem). From this point of view, our results have to be taken as another step on the way towards fully realistic numerical models of GCs hosting an IMBH. Beside predicting properties of such systems, our results may also serve for calibration of other approaches, e.g. semi-analytic calculations \citep*{fragk18,fragleiginkoc18}, scattering of individual binaries on the IMBH in isolated three-body approximation \citep{frgual18b} or Fokker-Planck methods of integration of star clusters. We finally stress that our modelisation of the high-velocity stars ejected from GCs is only a proxy to the real population. As discussed, the main properties of the ejected stars may be inferred from Eqs.~(\ref{eqn:vej}) -- (\ref{eqn:rateimbh}), but further studies that consider different cluster and IMBH sizes and binary properties are highly desirable to precisely model the ejected population of stars and compare it to other scenarios \citep{frgual18b}.

The mass of the cluster considered in our simulations together with an assumption of the canonical IMF implies less than 100 stellar remnants to be stellar-mass black holes. Despite the mass function of newly born stellar black holes is unknown as it is a subject to many processes (the stellar collapse itself, subsequent growth through mergers, ejection due to natal kicks and mutual scattering), we may expect that there may be no more than 10 black holes of mass $\gtrsim 10\,\msun$. This number is small enough to assume that this component of the cluster will not affect considerably its overall distribution. Nevertheless, it is likely that these massive stellar black holes will sink towards the IMBH and will dominate the central cusp (which is formed by no more than several tens of stars in our models {\MODBHBIN} and {\MODBH}). Hence, it is worth to discuss whether these massive black holes may influence the processes that take place in the vicinity of the IMBH, namely the Hill's process. In order to shed some light onto this problem, we have evaluated properties of orbits of binaries before one of their components was accelerated to high velocity. It appears that semi-major axes of these systems lie mostly in the range $\sim 0.1\,\pc$-$1\,\pc$, i.e. they are just penetrating into the cusp on radial orbits with high velocities and they spend only short time there. It is then likely that there will be a rather small probability of strong interaction with massive black holes and, therefore, we suggest that our results regarding ejection of stars from the cluster via interactions with the IMBH will not be considerably affected. What may be affected is observability of the central cusp which, if dominated by black holes, will not be detectable by conventional observations. (Note, however, that the cusp formed by several tens of stars will be definitely hard to observe anyway.)

We did not included an external tidal field, but we do not expect any significant changes regarding the processes in the vicinity of the IMBH, i.e. the formation of the cusp and the production of high velocity escapers. Only the presence of a tidal field would increase the evaporation of the cluster and the escape rate of stars with small velocities ($\lesssim 30 \kms$).

\section{Acknowledgements}
We thank Sverre J. Aarseth for helpful discussions on optimizations of the \code{NBODY6} code. This work was supported by the Czech Science Foundation through the project of Excellence No. 14-37086G. Most of the numerical calculations were performed on the computational cluster Tiger at the Astronomical Institute of the Charles University. GF acknowledges hospitality from Ladislav \v{S}ubr and Charles University of Prague, where the early plan for this work was conceived. GF is supported by the Foreign Postdoctoral Fellowship Program of the Israel Academy of Sciences and Humanities. GF also acknowledges support from an Arskin postdoctoral fellowship and Lady Davis Fellowship Trust at the Hebrew University of Jerusalem. 

\bibliographystyle{mn2e}
\bibliography{refs}
\end{document}